\documentclass[%
prd,
reprint,
superscriptaddress,
 amsmath,amssymb,
 aps,
]{revtex4-1}

\pdfoutput=1

\usepackage[utf8]{inputenc}

\usepackage{xcolor}

\usepackage{graphicx}% Include figure files
\usepackage{dcolumn}% Align table columns on decimal point
\usepackage{bm}% bold math

\newcommand{\rf}[1]{Eq.~(\ref{#1})}
\newcommand{\rfn}[1]{~(\ref{#1})}

\newcommand{\be}{\begin{eqnarray}}
\newcommand{\ee}{\end{eqnarray}}

\newcommand{\p}{\partial}

\newcommand{\f}[2]{\frac{#1}{#2}}

\newcommand{\sym}{${\mathcal N}=4$}

\newcommand{\pL}{{\mathcal P}_{||}}
\newcommand{\pT}{{\mathcal P}_{\perp}}
\newcommand{\edens}{{\mathcal E}}
\newcommand{\pa}{{\mathcal A}}
\newcommand{\pas}{{{\mathcal A}_\star}}
\newcommand{\paz}{{{\mathcal A}_\star}}
\newcommand{\dndy}{\f{dN}{dy}}
\newcommand{\dndyh}{dN/dy}

\begin{document}

\author{Jakub Jankowski\thanks{Jakub.Jankowski@fuw.edu.pl}}
\affiliation{Faculty of Physics, University of Warsaw,
%ul. Pasteura 5,
02-093 Warsaw, Poland}
\affiliation{Institute of Theoretical Physics, University of Wroc{\l}aw, 50-204 Wroc{\l}aw,
Poland}

\author{Syo Kamata}
\affiliation{National Centre for Nuclear Research, 02-093 Warsaw, Poland}

\author{Mauricio Martinez\thanks{mmarti11@ncsu.edu}}
\affiliation{Department of Physics, North Carolina State University, Raleigh, NC 27695, USA}

\author{Micha\l\ Spali\'nski}
% \email{michal.spalinski@ncbj.gov.pl}
\affiliation{National Centre for Nuclear Research, 02-093 Warsaw, Poland}
\affiliation{Physics Department, University of Bia{\l}ystok,
  % Konstantego Cio\l kowskiego 1L,
  15-245 Bia\l ystok, Poland}

\title{ Constraining the initial stages of ultrarelativistic nuclear collisions}

\begin{abstract} 

  It is frequently supposed that quark-gluon plasma created in heavy-ion
  collisions undergoes free streaming at early times. We examine this issue
  based on the assumption that a universal attractor dominates the dynamics
  already at the earliest stages, which offers a way to connect the initial
  state with the start of the  hydrodynamic expansion in an approximate but
  conceptually transparent fashion.  We demonstrate that the centrality
  dependence of the measured particle multiplicities can be used to
  quantitatively constrain the pressure anisotropy and find that it strongly
  depends on the model of the initial energy deposition. As an illustration, we
  compare three initial state models and show that they predict rather different
  early-time values of the pressure anisotropy. This suggests that while
  assuming free streaming prior to hydrodynamization may be compatible with some
  initial state models, in general features of the prehydrodynamic flow need to
  be matched with the model of the initial state.

\end{abstract}

\maketitle

\section{Introduction} 

Experimental studies of the dynamics of quark-gluon plasma (QGP) created  in
high energy nuclear collisions aim at understanding initial states over which
there is little direct control. The standard approach to modeling heavy-ion
collisions involves formation of QGP followed by nonequilibrium evolution until
proper-time of about 1 fm/$c$ and subsequently by hydrodynamic evolution until
the local temperature drops below the confinement scale and hadrons are formed.
After a stage of hadronic cascade the final particle distributions are measured.

Early phases of this process are usually assumed to be invariant under
longitudinal boosts~\cite{Bjorken:1982qr}. This approximation holds best for
central events at midrapidity. When this assumption is combined with rotational
and translational invariance in the plane transverse to the collision axis it
implies that observables of the system depend only on the proper time elapsed
after the collision.  Despite their limitations, these approximations open the door to
manageable, semianalytic considerations at all stages of QGP evolution. 
The best known consequence of this line of reasoning is the asymptotic late-time behaviour of
the energy density 
\begin{equation} \label{eq.bjorken} \edens \sim
\f{\Lambda^4}{(\Lambda\tau)^{4/3}}~, \end{equation}
where the scale $\Lambda$ is the only remnant of the initial conditions.  Since
at such late times the system is already close to local equilibrium, one can
translate this into a statement about the entropy density which can be directly
connected to the multiplicity of observed
hadrons~\cite{Bjorken:1982qr,Gyulassy:1983ub}.  The challenge is to relate such
physical observables to characteristics of the initial state.

In a series of recent developments it was realized that at
least in the case of boost invariant flow it may be reasonable to assume that
the evolving plasma system behaves in a predictable way already at the earliest
stages.  In some simple models of equilibration it was observed that certain observables
exhibit universal behavior very early on, not only at late times, when the
system is very close to equilibrium~\cite{Heller:2015dha,Aniceto:2015mto}.  Such
far-from-equilibrium attractors were later identified also in other models of
nonequilibrium dynamics such as strongly-coupled \sym\ supersymmetric Yang-Mills theory through
the AdS/CFT correspondence~\cite{Romatschke:2017vte,Spalinski:2017mel} and
weakly coupled models in kinetic
theory~\cite{Strickland:2017kux,Denicol:2017lxn,Mazeliauskas:2018yef,Strickland:2018ayk,Denicol:2019lio,Kurkela:2019set,Almaalol:2020rnu,Behtash:2019txb,Behtash:2020vqk}.

It has been argued recently that such early-time, far-from-equilibrium
attractors exhibit two distinct stages, both occurring at large values of
the pressure anisotropy which precludes a purely hydrodynamic interpretation in
the traditional sense. The later stage is determined by the decay of
nonhydrodynamic modes and its details depend on the microscopic
theory~\cite{Janik:2006ft,Heller:2007qt,Heller:2016rtz,Florkowski:2016zsi,Aniceto:2018uik}.
It is natural to expect that this type of behaviour should be generic and
independent of any special symmetry assumptions.  However, the earliest stage
which one observes in boost-invariant models appears to be dominated by the
rapid expansion successfully competing with nonhydrodynamic mode
decay~\cite{Blaizot:2017ucy,Kurkela:2019set,Heller:2020anv}.  The basic
observation underlying our work is that if this is the case, then attractor
behavior should also occur in QCD, since it is at least partly a kinematic
effect -- a consequence of approximate boost invariance.  This attractor would
then provide a bridge between the initial state and the start of hydrodynamic
evolution. This crucial stage, which sets initial conditions for hydrodynamics,
is often taken to be free-streaming.  The question we address is whether this
behavior is compatible with any given initial state model. 

The existence of an attractor implies a specific power-law dependence of the
energy density $\edens(\tau)$ for asymptotically small proper time $\tau$. For
example, in models based on kinetic theory one finds $\edens\sim 1/\tau$, which
corresponds to free streaming: the system expands starting from an initial state
where the longitudinal pressure vanishes~\cite{Baym:1984np}. This behaviour has
been explored in a number of phenomenological
studies~\cite{Liu:2015nwa,Kolb:2000sd,Broniowski:2008qk,JETSCAPE:2020mzn}.
However, the existence of a nonequilibrium attractor does not in itself imply
free streaming at early times.  For instance, in hydrodynamic models of
equilibration~\cite{Heller:2015dha,Aniceto:2015mto}, the early time behaviour of
the system is determined by the transport coefficients and can therefore be
tuned to the extent that those parameters can be varied.  

At the level of a microscopic theory such as QCD this early time behaviour is
not known, so we parameterise it as $\edens\sim\tau^{-\beta}$ in terms of a
constant parameter $\beta$.  The main technical point, presented in the
following Section, is that if an attractor exists, then the evolution of the
energy density can be determined up to a single integration constant, for any
value of the scaling exponent $\beta$.  The significance of this is that it
provides important information about the prehydrodynamic evolution of
quark-gluon plasma, as we will discuss below. 

The phenomenological analysis of heavy-ion collisions typically involves a model
of the initial state which supplies the initial conditions for hydrodynamic
evolution. More specifically, the nuclear thickness functions $T_A, T_B$ of the
colliding nuclei are generated~\cite{Miller:2007ri} and these are combined
(using some phenomenological model of the initial state) to produce the
energy-density profile at some very early proper-time $\tau_0$. This is then
evolved using free streaming, which results in a state of the energy-momentum
tensor suitable for initializing hydrodynamics. Our point of departure is the
observation is that instead of using free-streaming for this prehydrodynamic
stage of evolution one should instead try to mimic evolution along the
nonequilibrium attractor.  

It could be that the relevant attractor is actually free-streaming at very early
times~\cite{Giacalone:2019ldn}, but a priori this is not obvious. In this paper
we point out that the prehydrodynamic evolution is not independent of the choice
made for the model of the initial state. If an attractor governs early-time
dynamics, then given an initial state model the centrality dependence of the
expected particle multiplicities turns out to be quite sensitive to the
parameter $\beta$. In this way one can link an initial-state model with the
corresponding early-time behaviour of the energy density, which is equivalent to
identifying the early-time behaviour of the boost-invariant attractor. 

We find that the compatibility of free streaming with any given initial state
model cannot be assumed {\em a priori}.  In this exploratory study we consider
three models of the initial energy deposition involving various scalings. Our
goal however is not to study their relative merits, but to see whether they
favour different early-time asymptotics of the energy density, and thus
different attractors. In order to do this, in each case we calculate the value
of the parameter $\beta$ which leads to the best fit for the observed
multiplicities. While a very good fit can be found for each of these three
models, the actual value of $\beta$ obtained this way varies considerably.  This
shows that the prehydrodynamic stage of QGP evolution need not be well
approximated by free streaming, and the appropriate attractor depends on the
chosen initial state model.

\section{Attractor behaviour} 

In this section we describe how to exploit
attractor behaviour of the pressure anisotropy to approximate the full dynamics
of boost-invariant systems. We would like to emphasise that the analysis
presented here does not assume any specific features of the attractor.  In
particular, it does {\it not} assume free streaming at early times. 

In the case of Bjorken flow the expectation value of the energy-momentum tensor
can be parameterised as
\be 
\langle T^{\mu}_{{\ }\nu}\rangle = 
\mathrm{diag}(- \edens, \pL, \pT, \pT)^\mu_{{\ }\nu}~.  
\ee
We will focus on conformal systems, for which the energy-momentum tensor is
traceless and
\be 
\label{eq.plpt} \pL = \f{1}{3}\edens \left(1-\f{2}{3}\pa\right),\qquad \pT =
\f{1}{3}\edens \left(1+\f{1}{3}\pa\right)~, 
\ee
where $\edens$ is the energy density and $\pa$ is the {\em pressure anisotropy},
which is a dimensionless measure of how far the system is from local thermal
equilibrium.

The conservation of the energy-momentum tensor can be expressed in the form
\be 
\label{eq.cons} 
\tau\p_\tau \log \edens = -\f{4}{3} + \f{2}{9}\pa~.  
\ee
It will be convenient to define the effective temperature $T\equiv\edens^{1/4}$
up to a constant factor which will play no role in our considerations. After
introducing the dimensionless variable $w\equiv \tau T$, the conservation
equation~\eqref{eq.cons} is rewritten as
\be 
\label{eq.consw} 
\f{d\log T}{d \log w} = \f{\pa - \ 6}{\pa + 12}~.  
\ee
For a perfect fluid, $\pa=0$ and either \rf{eq.cons} or \rf{eq.consw} suffices to
determine the solution, leading to \rf{eq.bjorken}. However, for dissipative
systems one must also specify $\pa(w)$, which depends on the 
microscopic dynamics of the plasma as well as the initial state of the
system.

If $\pa(w)$ is given, one can integrate \rf{eq.consw} to solve for the effective temperature
as a function of $w$:
\begin{equation}
  \label{eq.evol}
   T(w) =  \Phi_\pa (w, w_0)  T(w_0)~,
\end{equation}
for some initial condition (integration constant) $T(w_0)$.
The function $\Phi_\pa$ reads~\cite{Giacalone:2019ldn}
\begin{equation}
  \label{eq.phidef}
  \Phi_\pa (w, w_0) = \exp\left(\int_{w_0}^{w} \f{dx}{x}
  \f{\pa(x) - \ 6}{\pa(x) + 12} \right)~.
\end{equation}
The subscript $\pa$ which appears above indicates the dependence of this
quantity on the pressure anisotropy as a function of $w$.

Although \rf{eq.evol} expresses the content of the conservation of
energy-momentum in the Bjorken setting, one still needs to determine $\pa(w)$
for a given solution, which will in general depend on additional information
characterizing the initial state. Crucially, in some model systems, such as
hydrodynamic models of equilibration~\cite{Heller:2015dha,Aniceto:2015mto} or
kinetic
theory~\cite{Blaizot:2017ucy,Romatschke:2017vte,Strickland:2017kux,Strickland:2018ayk,Kurkela:2019set,Blaizot:2020gql,Almaalol:2020rnu,Behtash:2019txb,Behtash:2020vqk,Kamata:2020mka}
there is now a lot of evidence pointing to universal behaviour of the pressure
anisotropy setting in very early on, when the system is still very far from
equilibrium. 
By this we mean that for a given range of initial conditions,
apart from an initial transient, the function $\pa(w)$ quickly approaches a
universal attractor $\pa_\star(w)$ which is determined by the
microscopic theory under consideration (perhaps numerically, or through some
sort of a ``slow roll'' approximation~\cite{Heller:2015dha}). We
assume that the physically interesting range of initial conditions is in the
basin of attraction of this unique attractor.  This suggests that it should be a
good approximation to replace the form of the pressure anisotropy $\pa(w)$, as
it appears in \rf{eq.consw}, by the attractor $\pa_\star(w)$:
\begin{equation}
  \label{eq.approx}
   T(w) \approx  \Phi_{\pa_\star} (w, w_0)  T(w_0)~.
\end{equation}
Within such an approximation, the temperature at late times is determined by the
temperature at early times {\em alone}: the remaining dependence on the initial
state is neglected by assuming that the effective dynamics of the system is
captured by its attractor, apart from a negligible initial transient.

The importance of \rf{eq.approx} rests on the fact that it is an explicit
relation between the initial and final states of expanding plasma.  It implies a
relationship between the initial state energy density and the entropy of the 
near-equilibrium system at late times. One can then estimate the multiplicities of
observed hadrons by following essentially the same method as outlined in
\cite{Giacalone:2019ldn} for the special case of a free-streaming attractor.

\section{Early time behaviour} 

As currently understood,  the existence of an
universal attractor is contingent upon there being a definite, finite,
physically distinguished behaviour of the pressure anisotropy at
$w=0$~\cite{Heller:2015dha,Aniceto:2015mto}. One can translate this into a
statement about the behaviour of the temperature at early times. Indeed, under
the above assumptions, the conservation of energy-momentum \rf{eq.cons} implies
that for asymptotically small proper-time $\tau$
\be
\label{eq.mu}
  \edens \sim \f{\mu^4}{(\mu\tau)^{\beta}}~.
\ee
We will focus on $0\leq\beta < 4$. The scale $\mu$ is an integration constant which 
reflects the initial conditions, and the exponent $\beta$ is related to the attractor by
\be
\label{eq.abeta}
\paz(0) = 6\left(1-\f{3}{4}\beta\right)~.
\ee
While different initial conditions will correspond to different values of the
scale $\mu$, the parameter $\beta$ characterises the attractor itself  and is
therefore a feature of the particular microscopic theory under consideration.

For instance, in M\"uller-Israel-Stewart theory the attractor is the unique
stable solution which is regular at $w=0$, where 
\be \label{eq.mis} \paz(0) = 6\sqrt{\frac{C_\eta}{C_{\tau\Pi}}} \iff \beta =
\f{4}{3} \left(1 - \sqrt{\f{C_\eta}{C_{\tau\Pi}}}\right)~,
\ee
and $C_\eta$, $C_{\tau\Pi}$ are dimensionless constants given by rescaling the
transport coefficients by appropriate powers of $T$
\cite{Heller:2015dha,Florkowski:2017olj}. 
Thus, in such cases the value of $\beta$ is determined by the transport coefficients. 

Aside from simple models, little is known about the existence of attractors or
the early time asymptotics of the energy density captured by the parameter
$\beta$. However, as we will show, one can constrain the value of $\beta$ by
calculating certain hadronic observables sensitive to the initial state and
comparing with experiment.

\section{Entropy} 

The universal attractor determines both early and late-time 
behaviour of the system, and this fact makes it possible to relate final state entropy to characteristics
of the initial state.
To streamline the notation we will denote
the value of $w$ at very early proper time $\tau_0$  by $w_0$, and its value at
late times $\tau_\infty$
by $w_\infty$. 

Using \rf{eq.approx} one finds the key relation between 
the entropy density per unit rapidity at late time and the initial energy density
\be
\label{eq.entrofinal}
s(\tau_\infty) \tau_\infty= h(\beta) \left(\edens(\tau_0) \tau_0^{\beta}\right)^{\frac{2}{4-\beta}}~,
\ee
where
\be
\label{eq.hbeta}
h(\beta) = \f{4}{3} w_\infty w_0^{\f{2\beta}{\beta-4}} \Phi_\pas(w_\infty, w_0)^2~.
\ee
We emphasize that \rf{eq.entrofinal}
is a universal, albeit approximate, statement in any model which possesses an attractor in the sense
under consideration here.

An important point about the function in \rf{eq.hbeta} is that it is actually {\em
independent} of the specific values of $w_0\ll 1$ and $w_\infty\gg 1$ appearing
on the right-hand side of this equation. This ensues because the quantity in
\rf{eq.phidef} diverges for small $w_0$ and vanishes for large $w_\infty$
precisely in such a way that the dependence on the initial and final values of
$w$ drops out in the implied asymptotic limits, leaving a finite and nonzero
result.  This can be shown in general based on the asymptotic behaviours of the
pressure anisotropy. 

\section{Connection to experiment} 

The estimate of entropy density, 
\rf{eq.entrofinal}, can be translated into a statement about centrality
dependence of observed particle multiplicities. 
Given the entropy density in 
\rf{eq.entrofinal}, the charged particle multiplicity of a specific event can be
expressed as 
\be
\label{eq.dndy}
\dndy = A \tau_0^{\frac{2\beta}{4-\beta}} h(\beta) \int d^2\mathbf{x}_\perp
\edens(\tau_0, \mathbf{x}_\perp)^{\frac{2}{4-\beta}}~,
\ee
where $A$ is a constant whose value will not be relevant to
our considerations.
The new element here is allowing for a nontrivial dependence of the initial energy
density on the location in the plane transverse to the collision axis. This
brings in dependence on the impact parameter of a given event. The underlying assumptions
and applicability of this procedure are discussed in
Ref.~\cite{Giacalone:2019ldn}.  
We will then use formula (\ref{eq.dndy}) to estimate the expected multiplicity by
averaging over Monte Carlo generated events.  
For a given event, the calculation of the energy density requires a model of the
initial state. 
In this work we consider three such models, all of which are formulated in terms
of a nuclear thickness function $T(\mathbf{x}_\perp)$  
which is obtained 
by Monte Carlo sampling~\cite{Miller:2007ri}. 

%%%%%%%%%%%%%%%%%%%%%%%%%%%
% \begin{figure*}[tp]
\begin{figure}[ht]
\includegraphics[width=0.48\textwidth]{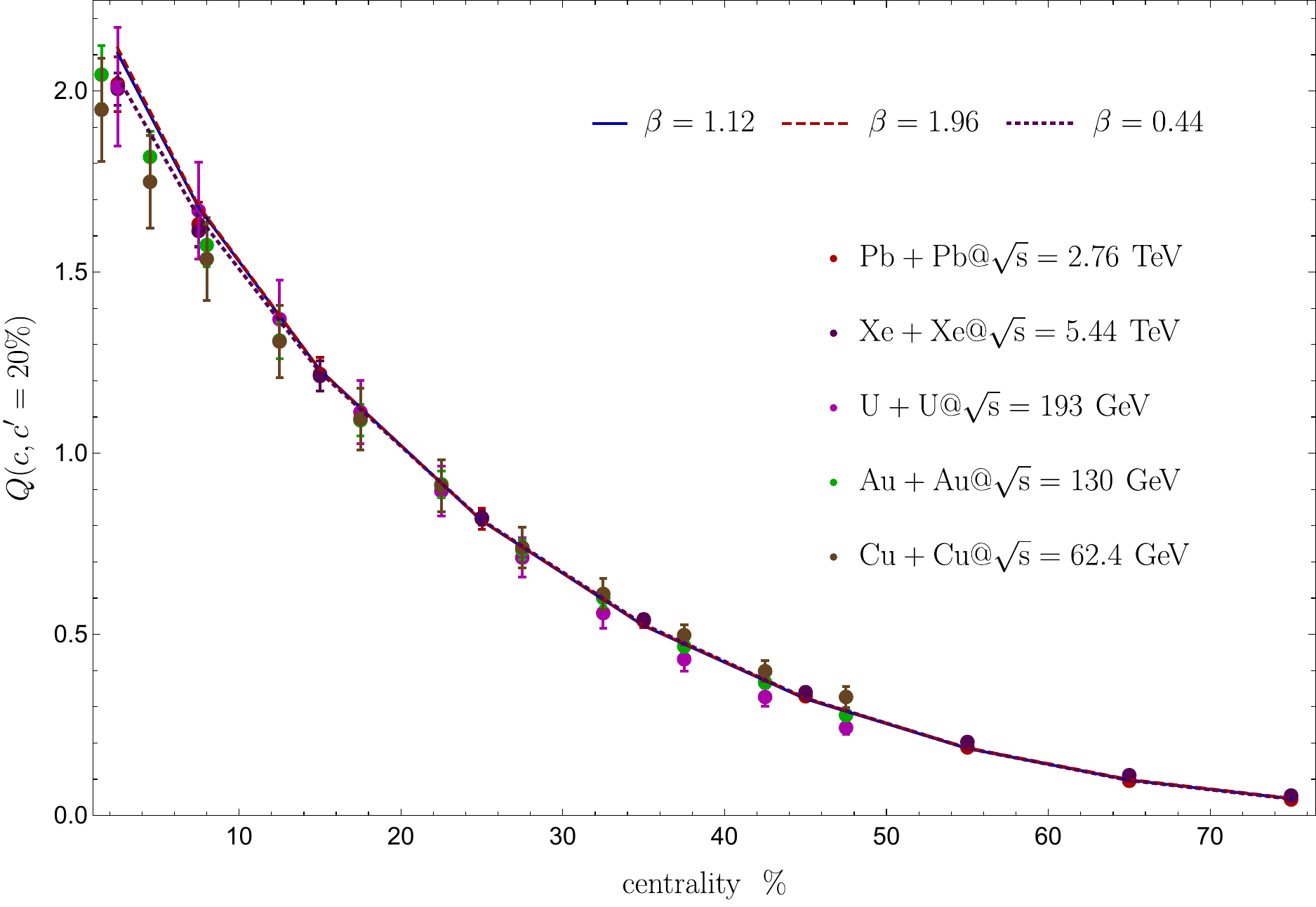}
\caption{
  \label{fig:wide} 
  Universal centrality dependence of $Q(c,c'{=}20\%)$, i.e. the number of
  produced charged particles normalized to $20\%$ centrality for each of the
  three models we consider. Experimental data
  shown for different collision systems: Xe+Xe \cite{Acharya:2018hhy}, Pb+Pb
  \cite{Aamodt:2010cz}, Au+Au \cite{Alver:2010ck},U+U \cite{Adare:2015bua},
  Cu+Cu \cite{Alver:2010ck}.
  }
\end{figure}
%%%%%%%%%%%%%%%%%%%%%%%%%%%

In the first model, which we will refer to as Model I, 
the energy density for a given event is given 
by~\cite{Dumitru:2001ux,Blaizot:2004wu,Gelis:2005pt,Blaizot:2010kh,Giacalone:2019ldn,Schlichting_2020}.
\be 
\label{eq.dd} 
\edens^{(I)} (\tau_0, \mathbf{x}_\perp) = C 
T^<(\mathbf{x}_\perp) \sqrt{T^>(\mathbf{x}_\perp)}
\ee
where the constant $C$ is independent of the impact
parameter $\mathbf{b}$, which enters only through 
\be
T^<(\mathbf{x}_\perp) =
\mathrm{min}(T(\mathbf{x}_\perp+\mathbf{b}/2),T(\mathbf{x}_\perp- \mathbf{b}/2))
\ee
with an analogous formula holding for $T^>(\mathbf{x}_\perp)$. 
The second model 
is the $p=-1$ case of the Trento family of
models~\cite{Moreland:2014oya}. In
this case 
\be 
\label{eq.harmonic} 
\edens^{(II)}(\tau_0, \mathbf{x}_\perp)  = C
\f{T(\mathbf{x}_\perp+\mathbf{b}/2) T(\mathbf{x}_\perp-\mathbf{b}/2)}{
T(\mathbf{x}_\perp+\mathbf{b}/2) + T(\mathbf{x}_\perp-\mathbf{b}/2)}
\ee
The third model is defined by~\cite{Lappi:2006hq,Romatschke:2017ejr}:
\be 
\label{eq.roma} 
\edens^{(III)} (\tau_0, \mathbf{x}_\perp) = C
T(\mathbf{x}_\perp+\mathbf{b}/2) T(\mathbf{x}_\perp-\mathbf{b}/2)
\ee
Here $C$ is again a normalization factor independent of the impact parameter. 

We now focus our discussion on the case of PbPb collisions at $\sqrt{s} = 2.76$
TeV at the LHC and confront our findings with the ALICE data, which quotes the
multiplicity of charged particles in each of $9$ centrality classes
\cite{Aamodt:2010cz}.  We will use our formula \rf{eq.dndy} to calculate the
corresponding prediction in each centrality class. The centrality $c$ is
connected to the impact parameter $b$ by the relation $c=\pi b^2/\sigma$, where
$\sigma=797~\mathrm{fm}^2$ is the total inelastic nucleus-nucleus cross-section
at $\sqrt{s} = 2.76$ TeV.

We define the following ratios of multiplicities at different centralities 
\be 
Q(c,c') \equiv \f{\langle\dndyh\rangle_{c^{\ }}}{\langle\dndyh\rangle_{c'}}~, 
\label{eq:Qij} 
\ee
where the angle-brackets denote the mean value over events in the specified
centrality class. These quantities are independent of the
normalization 
factors $C$ entering \rf{eq.dd}, \rfn{eq.harmonic}, \rfn{eq.roma}; they are also independent of the
factor $h(\beta)$, which contains the details of the presumptive attractor.
However, they retain dependence on the parameter $\beta$ itself, which is
related to the attractor by \rf{eq.abeta}. In this way, for any value of $\beta$, we obtain a set of
numbers $Q(c,c')$ which can be directly compared to 
% experiment.  The published experimental errors of
% the multiplicity measurements can be used to estimate errors of the quantities
% in \rf{eq:Qij} needed for the standard $\chi^2$ method which we use to fit
% $\beta$ for each of the three initial state models we consider.  
% The best fit in each case is found for 
published experimental results. 
The best fit for each of the three models is 
found to be 
\be
\beta^{(I)} = 1.12, \quad
\beta^{(II)} = 1.96, \quad
\beta^{(III)} = 0.44 .
\ee
with statistical errors not exceeding $0.02$. 
The corresponding longitudinal and transverse pressures are related to $\beta$ by 
\be
\pL &=& (-1 + \beta) \edens ,\\
\pT &=& (1- \beta/2) \edens .
\ee
The beta values for the three models differ by a factor of almost 4.5,
which is significant even though we are not attempting to account fully and
quantitatively for the errors arising from the various approximations we have
made. 
This shows that if indeed an attractor determines early time behaviour,
 it is strongly connected to the initial state model.

We can also normalize all the multiplicities to 20\% centrality -- this choice
gives a result closest to what is obtained from the fit to all data points. This
allows us to compare the prediction of our analysis to many independent
measurements.  In Fig. \ref{fig:wide} we show our prediction for the ratio
$Q(c,c'=20\%)$ for different collision systems as function of centrality,
normalized to $20\%$ centrality.  This is plotted together with data taken from
various
experiments~\cite{Acharya:2018hhy,Aamodt:2010cz,Alver:2010ck,Adare:2015bua}.

\section{Conclusions and outlook} 

An early-time attractor should be viewed as a bridge between an assumed initial
state model and the stage of hydrodynamical evolution.  By considering three
models of the initial energy deposition we have shown that the centrality
dependence of measured particle multiplicities can be used to constrain the
early-time behavior of the energy-momentum tensor implied by a given model of
the initial state. 

It is often assumed that irrespective of the initial state model, the
prehydrodynamic evolution of QGP can be approximated by free streaming, with the
longitudinal pressure vanishing at asymptotically small times.  In the three
models which we have considered here this is not the case. Instead, the
longitudinal pressure varies significantly between them.  Furthermore, none are 
really close to free streaming: even in the closest case the predicted pressure
anisotropy is about 30\% lower relative to what is expected for free streaming. 

In conclusion, we have argued that the existence of an early-time attractor
couples the pre-equilibrium evolution of the system to the initial state model.
This suggests, in particular, that depending on the model of the initial energy
deposition it may not be appropriate to assume a stage of free streaming
governing the interval between QGP formation and hydrodynamization. This
conclusion was reached using some simplifying assumptions and relies on the most
basic observables, but its impact could be even more significant when some
symmetry requirements are relaxed, so that more sophisticated observables can be
explored. 

Our study suggests that it would be worthwhile to look for other options for the
prehydrodynamic evolution. This should be particularly relevant for Bayesian
studies~\cite{Bernhard:2016tnd,Nijs:2020ors,Nijs:2020roc,JETSCAPE:2020mzn},
which scan over families of initial state models, but assume free-streaming
prehydrodynamic evolution for all of them.

\begin{acknowledgments}
We would like to thank T.~Altinoluk, W.~Florkowski,
M.P.~Heller, A.~Mazeliauskas,  S.~Mrówczyński, P.~Romatschke, T.~Schaefer,
S.~Schlichting and V.~Skokov for discussions and/or comments on the manuscript.
S.~K. and M.~S. are supported by the Polish National Science Centre Grant
2018/29/B/ST2/02457. J.~J.  was  supported by the Polish National Science Centre
(NCN) Grant 2016/23/D/ST2/03125. M.~M. is supported in part by the US Department
of Energy Grant No. DE-FG02-03ER41260 and BEST (Beam Energy Scan Theory) DOE
Topical Collaboration. 
\end{acknowledgments}

\bibliography{bibcon}
\bibliographystyle{bibstyl}

\end{document}